\documentclass{article}
\usepackage{arxiv}
\usepackage[utf8]{inputenc}
\usepackage[T1]{fontenc}
\usepackage{xcolor}
\usepackage{graphicx}
\usepackage{booktabs}
\usepackage{amsmath, amssymb}
\usepackage{lmodern}
\usepackage{calc}
\usepackage{verbatim}
\usepackage{etoolbox}
\usepackage[numbers,square,sort&compress]{natbib}

\definecolor{blendedblue}{rgb}{0.2, 0.2, 0.6}
\usepackage[bookmarks=true,
            hyperfootnotes=false,
            colorlinks=true,
            linkcolor=blendedblue,
            urlcolor=blendedblue,
            citecolor=blendedblue,
            linktocpage=false,
            hyperindex=true]{hyperref}
\usepackage{url}
\usepackage{doi}
\usepackage{orcidlink}

\usepackage[labelsep=period]{caption}

\renewcommand{\headeright}{}
\renewcommand\baselinestretch{1.4}

\usepackage{parskip}
\title{Environmental Effect Compensation in SHM through Conditional Covariances: Application to the Icebreaker ``Polarstern''}

\renewcommand{\shorttitle}{Environmental Effect Compensation in SHM through Conditional Covariances}

\date{}
\renewcommand{\undertitle}{}

\makeatletter
\patchcmd{\@maketitle}{\textsc{\undertitle}\\}{}{}{
  \PackageError{main}{Could not remove the preprint subtitle}{Check arxiv.sty.}}
\patchcmd{\@maketitle}
  {\vskip 0.4in \@minus 0.1in \center{\@date} \vskip 0.2in}
  {\par\vskip 0.2in}{}{
  \PackageError{main}{Could not remove the title date}{Check arxiv.sty.}}

\renewcommand{\footnoterule}{\kern-3pt\hrule width\textwidth\kern2.6pt}
\newcommand{\publicationnote}{%
  \begingroup
    \renewcommand{\thefootnote}{}%
    \long\def\@makefntext##1{\noindent##1}%
    \footnotetext{%
      \renewcommand{\baselinestretch}{1}\footnotesize
      \textbf{Published version:}
      Simon Schlumbohm, Lizzie Neumann, Philipp Wittenberg, and Jan Gertheiss (2026).
      Environmental Effect Compensation in SHM through Conditional Covariances:
      Application to the Icebreaker ``Polarstern''.
      In: \emph{Proceedings of the XIII International Conference on Structural Dynamics
      (EURODYN 2026)}, pp.~1389--1400.
      DOI: \href{https://doi.org/10.47964/1326.10750.061}{10.47964/1326.10750.061}.}%
  \endgroup
}
\makeatother

\author{
    Simon Schlumbohm \orcidlink{0000-0002-0083-5142} \\
    Chair of Statistics and Data Science \\
    Dept. of Mathematics and Statistics\\
    School of Business, Economics\\
    and Social Sciences\\
    Helmut Schmidt University\\
    Hamburg, Germany\\
    \texttt{schlumbs@hsu-hh.de} \\
    \And
    Lizzie Neumann \orcidlink{0000-0003-2256-1127} \\
    Chair of Statistics and Data Science \\
    Dept. of Mathematics and Statistics\\
    School of Business, Economics\\
    and Social Sciences\\
    Helmut Schmidt University\\
    Hamburg, Germany\\
    \texttt{neumannl@hsu-hh.de} \\
    \AND
    Philipp Wittenberg \orcidlink{0000-0001-7151-8243} \\
    Chair of Statistics and Data Science \\
    Dept. of Mathematics and Statistics\\
    School of Business, Economics\\
    and Social Sciences\\
    Helmut Schmidt University\\
    Hamburg, Germany\\
    \texttt{pwitten@hsu-hh.de} \\
    \And
    Jan Gertheiss \orcidlink{0000-0001-6777-4746} \\
    Chair of Statistics and Data Science \\
    Dept. of Mathematics and Statistics\\
    School of Business, Economics\\
    and Social Sciences\\
    Helmut Schmidt University\\
    Hamburg, Germany\\
    \texttt{gertheij@hsu-hh.de} \\
}

\begin{document}
\maketitle
\publicationnote

\begin{abstract}
In structural health monitoring (SHM) systems, data is collected from a multitude of sensors measuring, for example, vibration, the strain on the structure, and potentially multiple different, additional features holding environmental or operational information. Notably, changes in the measured output do not necessarily originate from structural damage but are often induced by environmental changes. Accurately determining these confounding factors -- and subsequently removing their influence -- is therefore of high importance for reliable diagnostics. Many current methods for compensating environmental or operational effects rely on various forms of so-called response surface modeling, which, however, overlook possible variations in higher-order statistical moments. To address this issue, changes in the (conditional) covariance can be analyzed and quantified using a suitable non-parametric, kernel-based estimator.

In prior work, this methodology has been successfully applied to multiple bridges and output quantities such as natural frequencies as well as strain or inclination data.
In this work, we consider data gathered during the 2019/2020 MOSAiC Arctic expedition, in which the icebreaker ``Polarstern'' was equipped with 36 strain sensors positioned at various points across the vessel's deck and frame, as well as ambient temperature sensors. It is shown that both the strain data themselves and the correlations between different strain measurements are strongly affected by temperature, highlighting its confounding effect and the potential to obscure diagnostically relevant signals, which cannot be easily explained or removed via response surface modeling alone.
\end{abstract}

\keywords{Temperature influence, Conditional covariance, Conditional principal component analysis, Structural health monitoring, Environmental and operational variability, Non-parametric estimation.}

\bigskip

\section{Introduction}

The automated assessment of damage in any engineered system is based on sensor readings, which monitor the structure, device, or vessel in question. This process is commonly known as structural health monitoring (SHM) ~\cite{Farrar2012}.
The sensors used can be of varying nature, be it measurement of the environment (e.g., ambient temperature, wind speed or air pressure) or strain, which describes the contraction or expansion of the material, usually described as the change in length ($\Delta L$) relative to the original length ($L$), yielding a dimensionless value $\varepsilon$ defined as:

\begin{equation}
\varepsilon = \frac{\Delta L}{L}.
\label{eq:strain}
\end{equation}

Modern SHM systems are built from large sensor networks, which continuously record changes in the structure and yield heterogeneous data. The fundamental objective of these systems is reliable detection of structural damage, inferred from changes in measurements at different time points. However, variations in a strain sensor's signal are not solely caused by, and thus not wholly explained by, such structural deterioration alone. In practice, so-called environmental and operational variability (EOV) frequently impacts measurement responses. Their impact is described as comparable or potentially larger than the impact caused by the actual damage (i.e., usually the effect of interest) ~\cite{Sohn2006, Keshmiry2023}. Temperature fluctuations, humidity, or operational states may thus significantly influence both the mean behavior and the variability of structural responses -- the presence of such confounding effects is one of the central challenges in SHM~\cite{Pezeshki2023, Rezazadeh2025, Farrar2025}.

If not properly accounted for, environmental influence may cloud damage-related signatures and lead to false-positive indications or reduce the probability of detecting actual damage. Consequently, substantial research effort has been devoted to compensating for EOV effects. A widely used approach is so-called response surface modeling, where the structural response is modeled as a function of measured environmental or operational variables ~\cite{Worden2016, Worden2018, Kamali2025}. While effective in various applications, these techniques primarily focus on changes in first-order statistics (particularly the conditional mean) and ignore potential variation in higher-order statistical moments.

From a statistical perspective, structural damage may manifest not only as a shift in the expected value of a response quantity but also as changes in its variance or in the correlation between multiple sensor signals. A comprehensive compensation framework should hence account for alterations in the full conditional distribution of the monitored quantities. For instance, suitably designed nonparametric, kernel-based estimators provide more flexible means to (also) analyze and quantify changes in conditional covariance structures. In previous studies, this methodology has been successfully applied to several structures, including prestressed concrete and steel–concrete composite
bridges~\cite{Neumann2025b, Neumann2025} and a steel railway bridge~\cite{Neumann2025}.

In this study, building upon this foundation, data acquired during the 2019/2020 MOSAiC Arctic expedition is used and investigated ~\cite{Kubiczek2022, Kubiczek2022a}. The expedition was carried out by the German icebreaker and research vessel ``Polarstern'' and included a total of 36 strain sensors (measuring $\varepsilon$, see equation \ref{eq:strain}) distributed across both the deck and the ship's frame. Furthermore, 16 sensors continuously measured the ambient temperature (in °C) every 300 seconds in the predominantly Arctic environment. The operation spanned roughly one year, from late 2019 until late 2020, during which the sensors generated continuous data for subsequent analysis.

\begin{figure}[hbt!]
    \centering
    \includegraphics[width=0.9\textwidth]{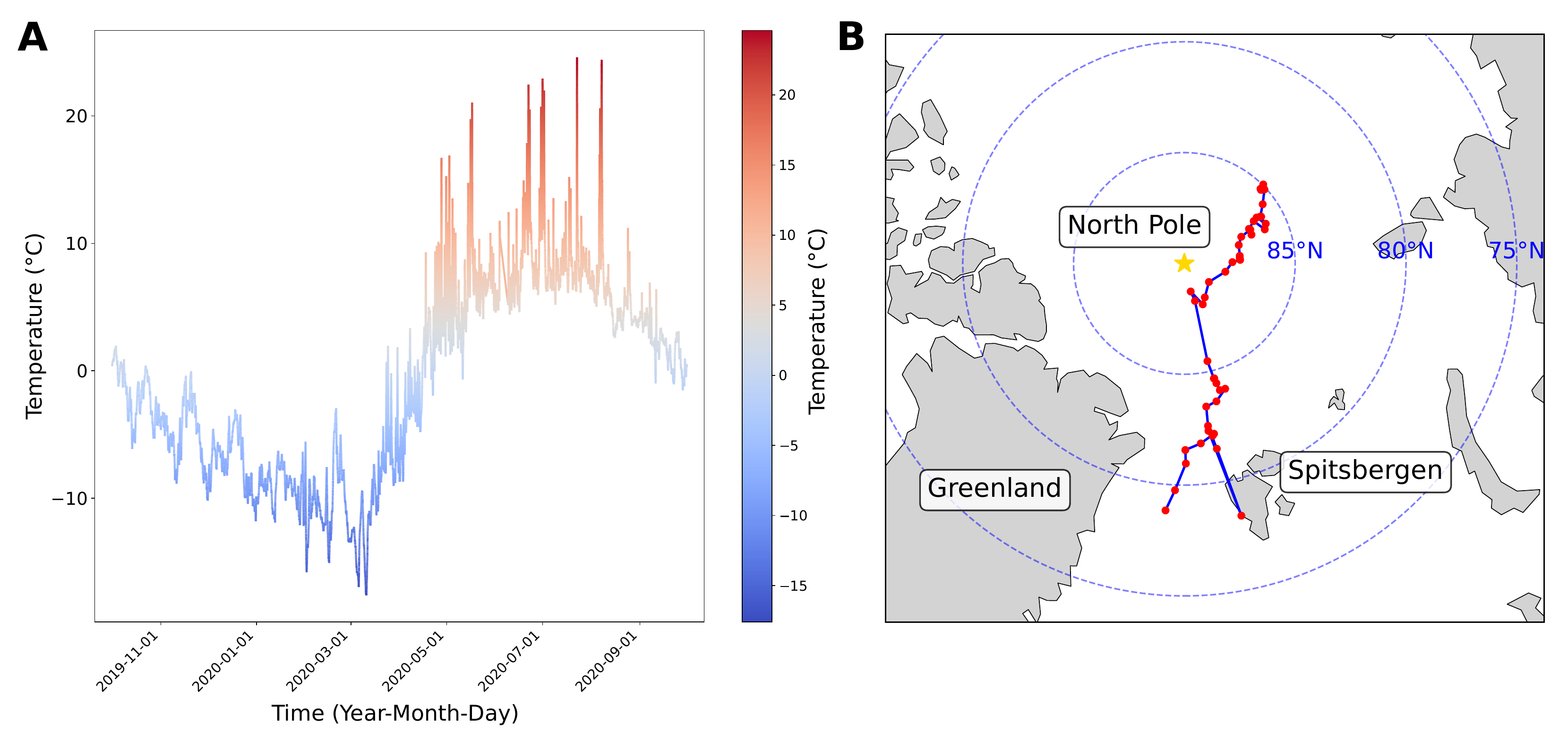}
    \caption{Overview of the temperature and the spatial trajectory during the MOSAiC expedition. \textbf{A} Average ambient temperature over time, averaged across available sensors (excluding those with insufficient data). \textbf{B} Ice drift trajectory of the ``Polarstern'', showing the predominantly southward movement within the Arctic ice, based on published data ~\cite{Kubiczek2022, Kubiczek2022a}.}
    \label{fig:MOSAiC_overview}
\end{figure}

The temperature over the course of the expedition is shown in Figure~\ref{fig:MOSAiC_overview}, panel~A. What is shown here is the average of all temperature sensors. A few sensors were excluded due to too few data points overall, possibly because of (partial) malfunction. 
The bulk of the MOSAiC expedition took place during a passive ice drift within the Arctic Circle, predominantly moving southwards. This drift is primarily shown in Figure~\ref{fig:MOSAiC_overview}, panel~B.
The coordinates on which this plot is based were obtained from the original publications ~\cite{Kubiczek2022, Kubiczek2022a}.

\begin{figure}[hbt!]
    \centering
    \includegraphics[width=0.95\textwidth]{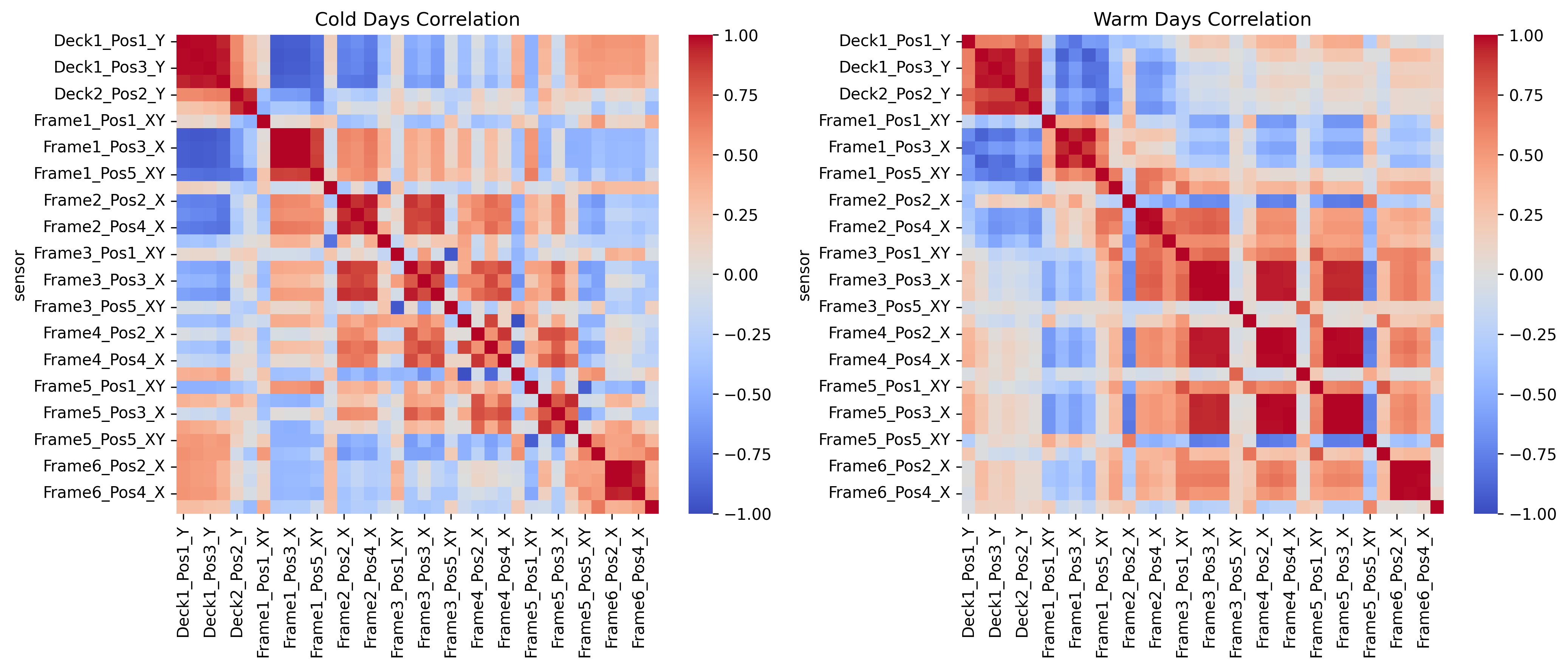}
    \caption{Correlation matrices visualized via heatmaps of the 36 strain sensors for two subsets of the data corresponding to the lowest and highest 25\% of average daily temperatures (termed "cold days" and "warm days"), respectively, prior to detrending.}
    \label{fig:cold_warm_before_detrend}
\end{figure}

In order to solidify the necessity of investigating variations in higher-order statistical moments, strain data from the ``Polarstern'' is analyzed with respect to the measured temperature on the respective expedition days. Two subsets are extracted from the data, each comprising 25\% of the total number of days: the quarter of days with the lowest average temperature and the quarter with the highest average temperature. For each subset, all 36 strain sensors are correlated against one another, yielding correlation matrices, which, when visualized as heatmaps, look as Figure~\ref{fig:cold_warm_before_detrend}.

As can be clearly seen, the temperature also alters the correlation between certain strain sensors, indicated by the differences between the two shown heatmaps. Commonly, detrending follows, with the goal to remove temperature influences. Here, detrending means fitting a linear model~(\ref{eq:linear_model}) for each (strain) sensor output $y_j$,

\begin{equation}
    y_j=\alpha_{j}+\beta_{j}z+e_j,
\label{eq:linear_model}
\end{equation}

with intercept $\alpha_{j}$, slope $\beta_{j}$, average temperature $z$, and error term $e_j$, $j=1,\ldots,36$. The residual is then defined as the difference between the observed strain and its fitted conditional mean, given the (average) temperature,

\begin{equation}
    \hat{e}_j=y_j-\hat{y}_j.
\end{equation}

\begin{figure}[hbt!]
    \centering
    \includegraphics[width=0.9\textwidth]{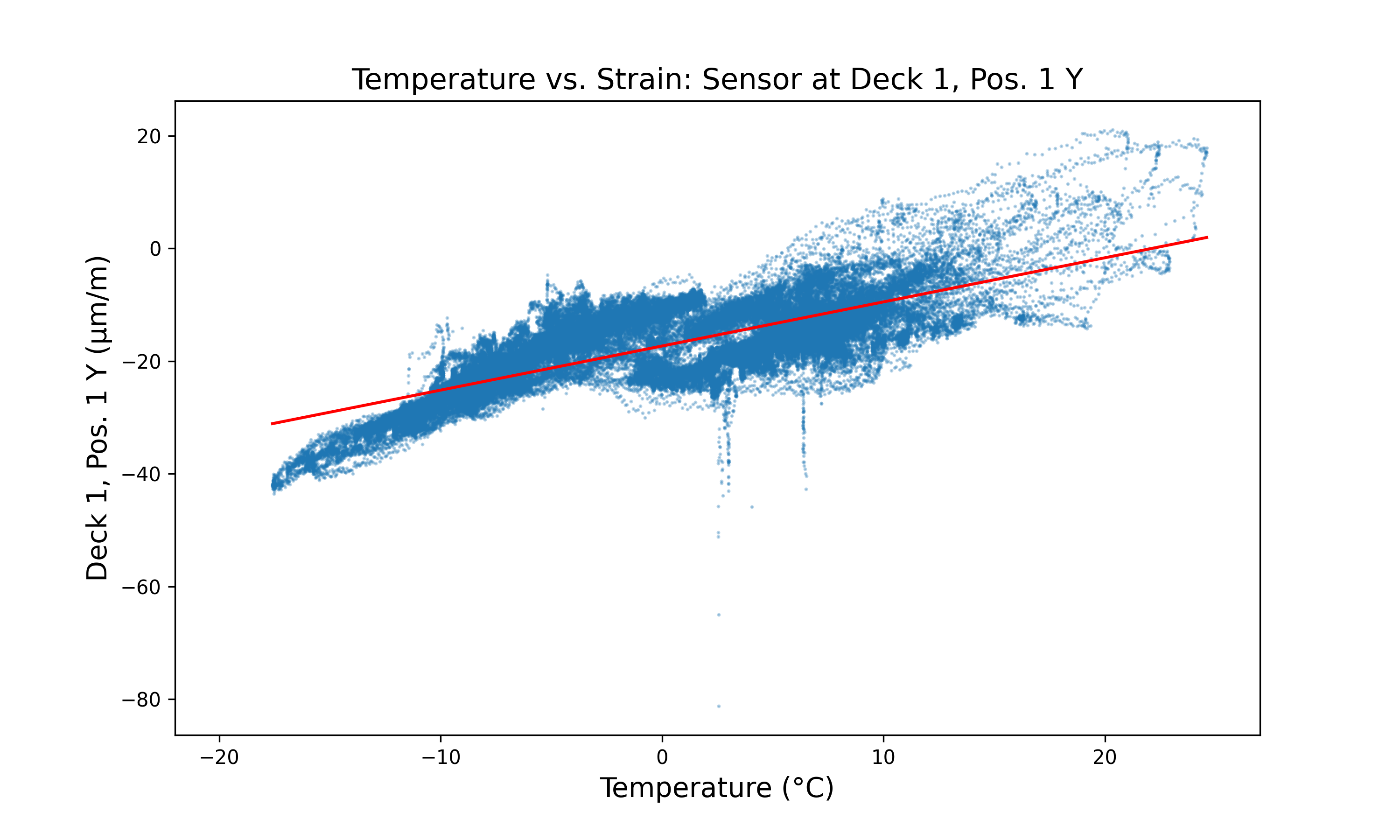}
    \caption{Detrending of a representative strain sensor on the ship’s deck. The fitted linear model captures the temperature-dependent mean behavior, while the residuals represent the detrended signal with first-order temperature effects removed.}
    \label{fig:Deck1_Pos1_Y_temp_strain}
\end{figure}

Detrending through linear regression is exemplarily shown for one of the deck sensors in Figure~\ref{fig:Deck1_Pos1_Y_temp_strain}.

\begin{figure}[hbt!]
    \centering
    \includegraphics[width=0.95\textwidth]{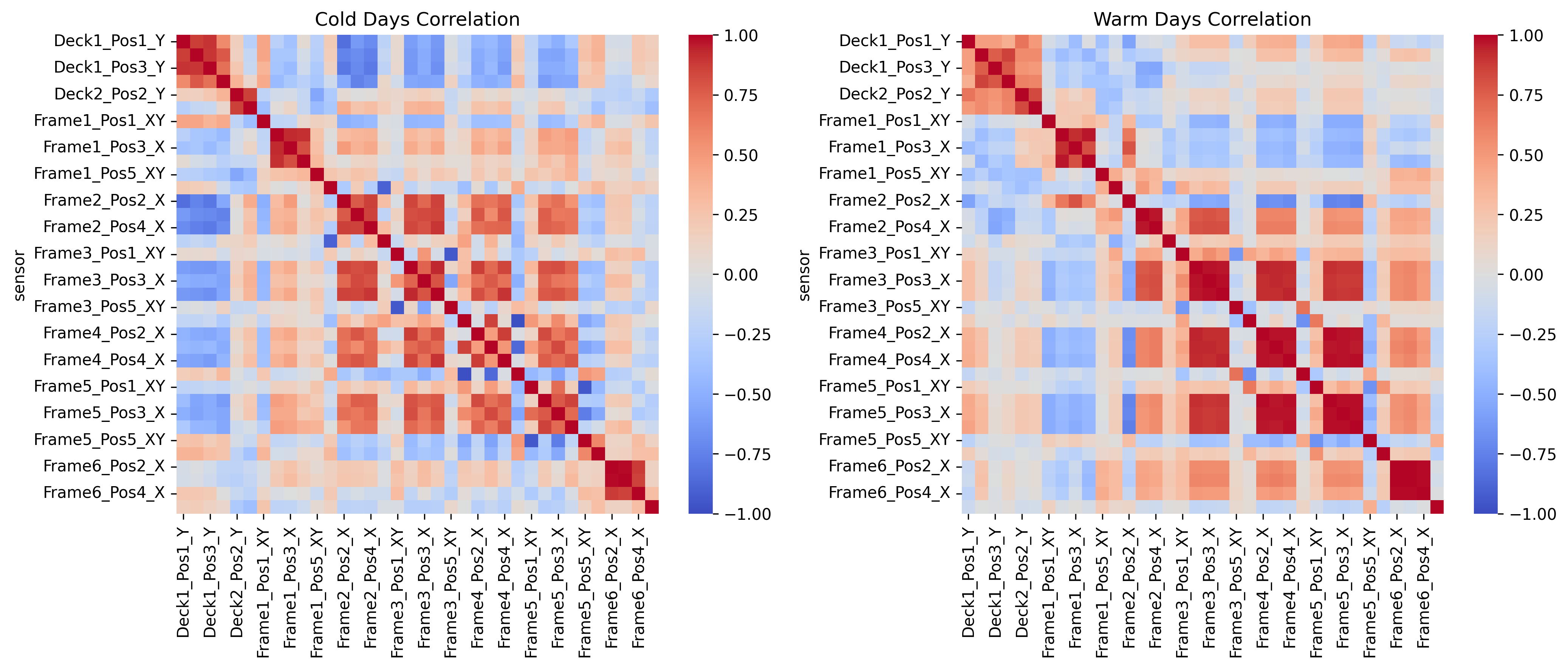}
    \caption{Correlation matrices visualized via heatmaps of the 36 strain sensors for two subsets of the data corresponding to the lowest and highest 25\% of average daily temperatures, respectively, after detrending.}
    \label{fig:cold_warm_after_detrend}
\end{figure}

Figure \ref{fig:cold_warm_after_detrend} shows the correlations, analogously to Figure~\ref{fig:cold_warm_before_detrend}, after detrending. It is visible that removing the mean dependence through detrending alone is insufficient, as temperature effects may still persist in the covariance structure (the heatmaps in Figure~\ref{fig:cold_warm_before_detrend} still look very similar to those from Figure~\ref{fig:cold_warm_after_detrend}). In other words, the dependency structure between sensors remains temperature-dependent even after removing mean effects.

This demonstrates that both the strain measurements themselves and the correlations between different strain sensors are strongly influenced by temperature, highlighting its pronounced confounding effect and underscoring the necessity of modeling changes in higher-order statistical moments as well, which conventional response surface modeling alone cannot adequately capture. The subsequent manuscript will be structured around this focus.

\section{Methodology}

This section introduces the statistical framework used to quantify and analyze temperature-dependent changes in the strain data from the Polarstern. The focus lies on modeling not only the conditional mean but also the conditional covariance structure of the multivariate sensor responses. To this end, a nonparametric estimator for the temperature-dependent covariance matrix is employed (Section~\ref{Non-parametric Conditional Covariance Matrix Estimation}), followed by a conditional principal component analysis (Section~\ref{Conditional Principal Component Analysis}) to extract interpretable features for further analysis. Both methodologies will then be employed for analyzing the MOSAiC expedition's data.

\subsection{Nonparametric Conditional Covariance Matrix Estimation}
\label{Non-parametric Conditional Covariance Matrix Estimation}

Let $\mathbf{y} = (y_1,\dots,y_p)^\top \in \mathbb{R}^p$ denote a $p$-dimensional random vector representing the considered outputs, and $z \in \mathbb{R}$ denotes a potential confounding variable. As an example, $\mathbf{y}$ may contain the natural frequencies associated with $p$ modes of a structure (e.g., the aforementioned bridges), or -- in the context of this work -- measurements of $p$ different strain sensors of a marine vessel, while $z$ may represent the ambient temperature (the confounder). Then, $\Sigma(z)$ is the conditional covariance matrix of $\mathbf{y}$ given $z$. A nonparametric estimator of this matrix can be constructed as follows:

\begin{equation}
\hat{\Sigma}(z;h)
=
\left\{
\sum_{i=1}^{n}
K_h(z_i - z)
\left[\mathbf{y}_i - \hat{\mathbf{m}}(z_i)\right]
\left[\mathbf{y}_i - \hat{\mathbf{m}}(z_i)\right]^\top
\right\}
\left\{
\sum_{i=1}^{n} K_h(z_i - z)
\right\}^{-1},
\label{eq:conditional_cov}
\end{equation}

where $\mathbf{y}_i = (y_{i1},\dots,y_{ip})^\top$, $i=1,\dots,n$, are observed samples of $\mathbf{y}$ and $z_i$ is the corresponding temperature ~\cite{Neumann2025, Yin2010}. The function $K_h(\cdot)$ denotes a kernel with bandwidth parameter $h$, and $\hat{\mathbf{m}}(z_i)$ represents an estimate of the conditional mean $\mathbf{m}(z)=\mathbb{E}[\mathbf{y} | z]$ evaluated at $z_i$; $\mathbf{m}(z)$ may be estimated by any suitable response surface modeling technique, e.g., a (multivariate) linear model~(\ref{eq:linear_model}) as shown in Figure~\ref{fig:Deck1_Pos1_Y_temp_strain} for one component, or by use of a nonlinear, kernel-based estimator~\cite{Yin2010,Neumann2025}. In applications involving structural vibration data, a piecewise linear regression model with a single breakpoint (often referred to as a bilinear model) is frequently employed ~\cite{Maes2022}. Important to note is that (\ref{eq:conditional_cov}) does not simply calculate the (empirical) conditional covariance on categorized (temperature) data as done in Figures~\ref{fig:cold_warm_before_detrend} and \ref{fig:cold_warm_after_detrend}, but allows for estimating the conditional covariance as a smooth function of an inherently continuous quantity (the temperature $z$).

\subsection{Conditional Principal Component Analysis}
\label{Conditional Principal Component Analysis}

In order to perform conditional principal component analysis (CPCA), an eigenvalue decomposition is applied to the conditional covariance matrix -- the previously introduced $\Sigma(z)$, as defined in \eqref{eq:conditional_cov}. Specifically~\cite{Neumann2025},

\begin{equation}
\Sigma(z) = A(z)\Lambda(z)A(z)^\top,
\end{equation}

where $\Lambda(z) = \mathrm{diag}(\lambda_1(z),\dots,\lambda_p(z))$ contains the eigenvalues ordered from largest to smallest. $A(z) = [\,a_1(z)\ \cdots\ a_p(z)\,]$ is the matrix whose columns are then the corresponding eigenvectors. These eigenvectors define the conditional principal component directions. As the covariance matrix depends on the confounder $z$, the eigenvalues and eigenvectors also vary with $z$. After estimating the eigenvalues, represented via $\hat{\lambda}_j(z)$, and estimating the eigenvectors, represented via $\hat{a}_j(z)$, they may serve to construct features for monitoring purposes. The associated vectors of conditional principal component scores are computed as follows~\cite{Neumann2025}:

\begin{equation}
\mathbf{s}_i
=
(\mathbf{y}_i - \hat{\mathbf{m}}(z_i))^\top
\hat{A}(z_i)
(\hat{\Lambda}(z_i))^{-1/2},
\label{eq:scores}
\end{equation}

where

\begin{equation}
(\hat{\Lambda}(z_i))^{-1/2}
=
\mathrm{diag}\big(\hat{\lambda}_1^{-1/2}(z_i),\dots,\hat{\lambda}_p^{-1/2}(z_i)\big),
\qquad
\hat{A}(z_i) = [\,\hat{a}_1(z_i)\ \cdots\ \hat{a}_p(z_i)\,].
\end{equation}

By centering with the conditional mean and scaling using the conditional eigenstructure, the resulting component scores are adjusted for the influence of the confounder $z$, meaning that the impact of the latter on the data used for potential/subsequent monitoring is removed.

\section{Results}

\begin{figure}[hbt!]
    \centering
    \includegraphics[width=1\textwidth]{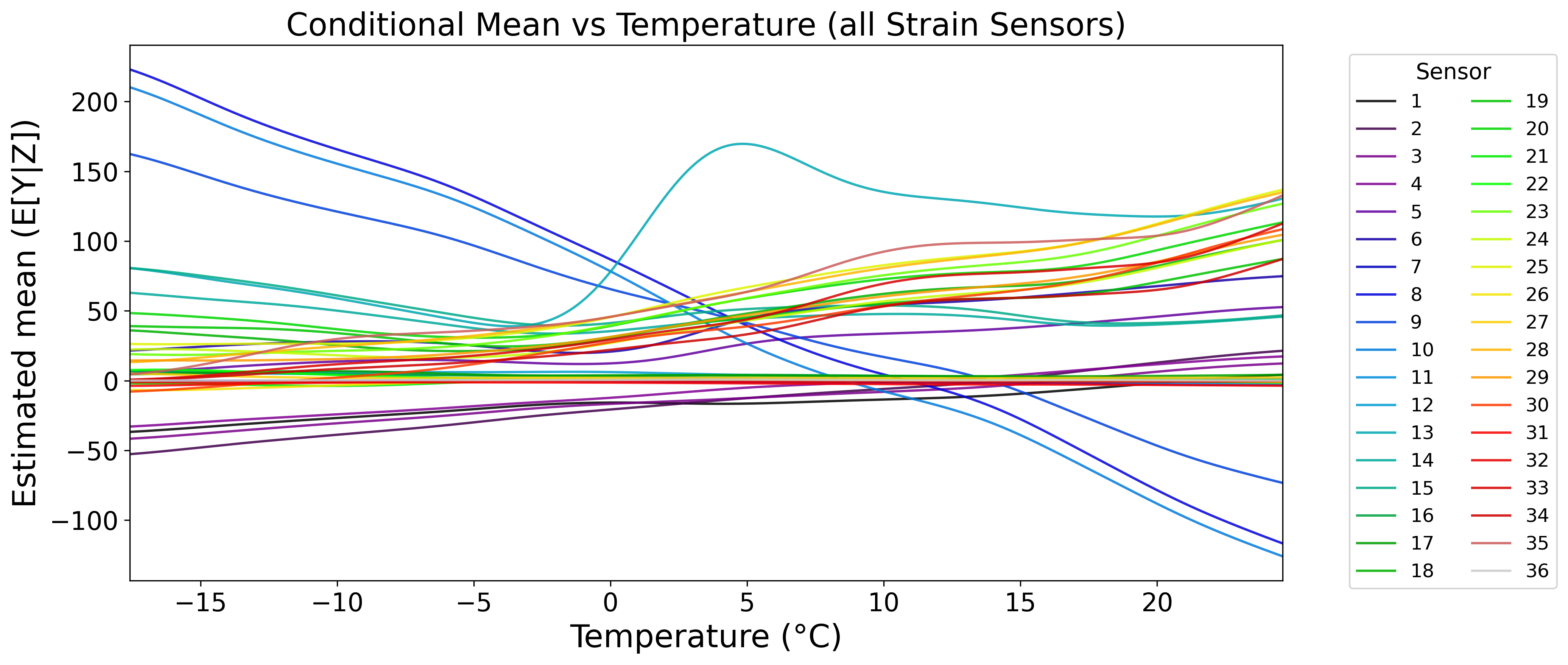}
    \caption{The estimates of the conditional mean $\mathbb{E}[y_j | z]$ for all strain sensors $j=1,\ldots,36$, visualized across the full temperature range.}
    \label{fig:estimated_means}
\end{figure}

The estimates of the conditional mean $\mathbb{E}[y_j | z]$ for each strain sensor $j=1,\ldots,36$ against the temperature $z$ in °C is depicted in Figure~\ref{fig:estimated_means}. It describes the typical strain response of each sensor at a given temperature and serves as a basic first-order characterization of the environmental influence. The estimates were obtained using a nonparametric, kernel-based approach that can identify and fit both linear and nonlinear relationships. As seen from Figure~\ref{fig:estimated_means}, most associations are (approximately) linear. So, a linear model like the one from Figure~\ref{fig:Deck1_Pos1_Y_temp_strain} would be suitable as well. 

\begin{figure}[hbt!]
    \centering
    \includegraphics[width=0.7\textwidth]{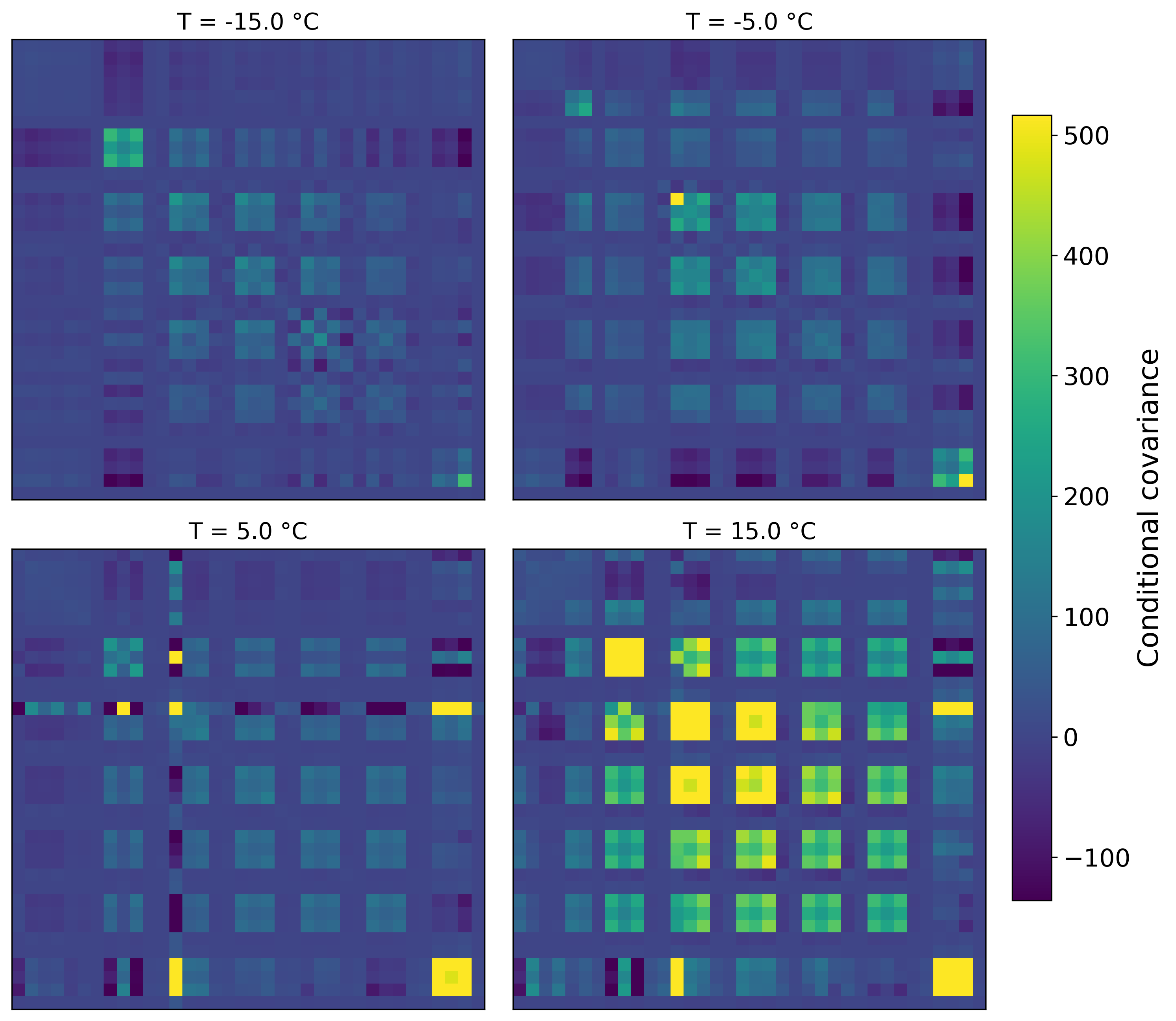}
    \caption{Heatmap representations of the estimated conditional covariance matrices of the 36 strain sensors. Each panel illustrates the pairwise covariance between sensor signals at a fixed temperature, with each pair equidistant, revealing how both the magnitude and structure of variability across the sensor network change with environmental conditions.}
    \label{fig:conditional_covariance_snapshots}
\end{figure}

This paper, however, focuses on investigating the effects beyond the mean behavior. To this end, temperature-dependent covariance matrices were estimated using the nonparametric approach described in Section~\ref{Non-parametric Conditional Covariance Matrix Estimation}. Figure~\ref{fig:conditional_covariance_snapshots} exemplarily shows the calculated conditional covariances for four specific, equidistant temperatures, visualized as heatmaps. The comparison of these heatmaps indicates that the covariance structure of the strain measurements varies systematically with temperature. In particular, both the variances (the diagonal) and the pattern of dependencies between sensors change across temperature levels, indicating that environmental conditions affect second-order statistical properties of the system.

\begin{figure}[hbt!]
    \centering
    \includegraphics[width=0.7\textwidth]{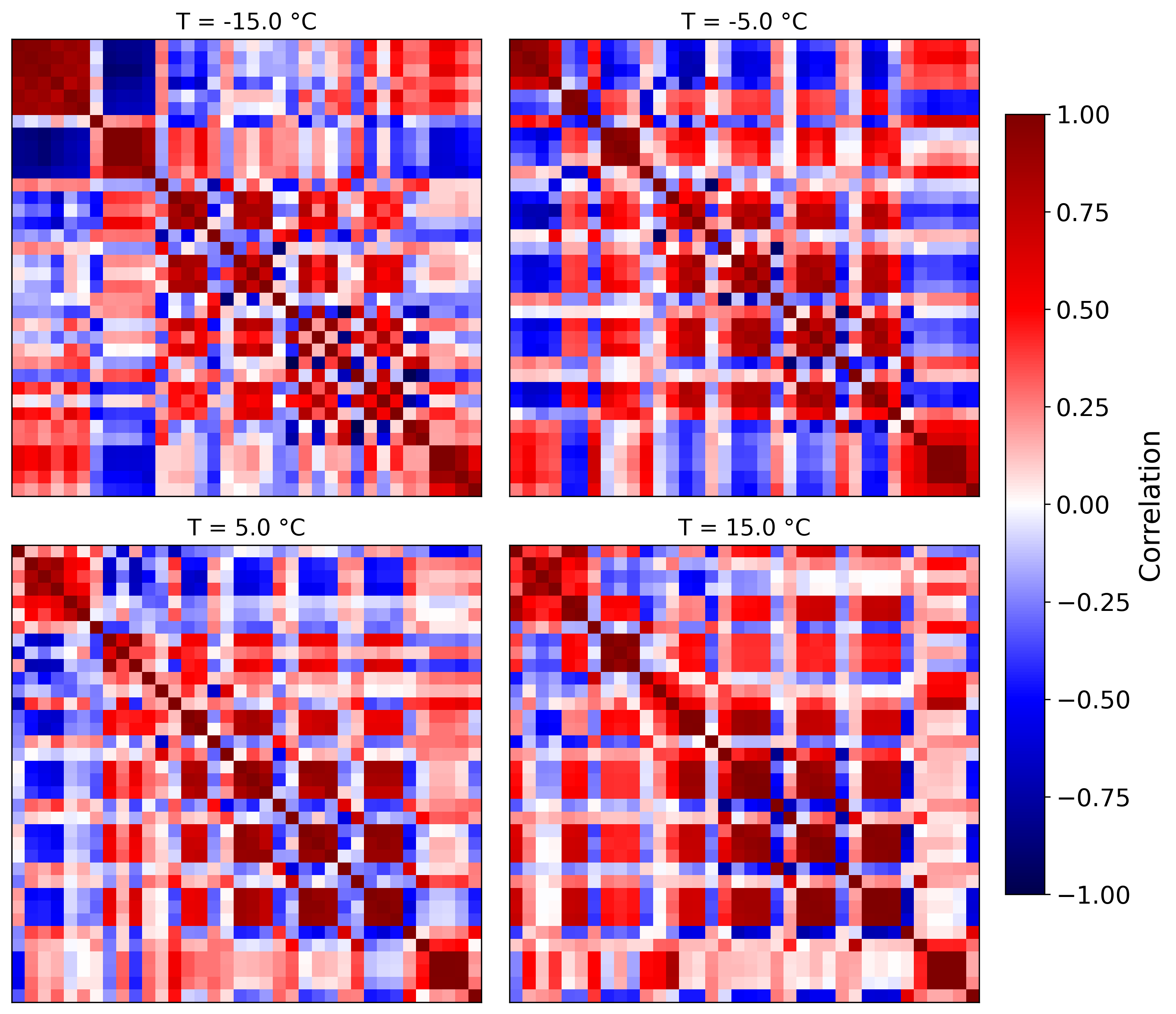}
    \caption{Heatmap representations of the correlation matrices derived from the temperature-dependent covariance estimates shown in Figure \ref{fig:conditional_covariance_snapshots}, evaluated at the same four equidistant temperature levels. By normalizing for sensor-specific variance, these plots highlight the underlying dependency structure between sensors.}
    \label{fig:correlation_snapshots.png}
\end{figure}

In order to isolate the pure dependency structure independent of absolute strain magnitudes and variances, the covariance matrices were subsequently transformed into correlation matrices. The resulting heatmaps mirror the results of Figure~\ref{fig:conditional_covariance_snapshots} in terms of sensor ordering and chosen temperatures, yet provide a normalized view of sensor coupling (Figure~\ref{fig:correlation_snapshots.png}). These visualizations enable a more intuitive interpretation and further emphasize that the inter-sensor relationships are not constant but exhibit pronounced temperature dependence. Distinct correlation patterns emerge at different temperature levels, demonstrating that the coupling between structural responses is substantially modulated by environmental conditions.

Overall, these findings confirm that temperature influences not only the mean strain response but also the covariance and correlation structure among sensors. This provides clear evidence that second-order statistical effects are present and relevant beyond the previously discussed, static bridge structures, as the data suggests the same issue applies to maritime vessels. Generally, this supports the need for the methodologies shown that explicitly account for changes in higher-order statistical moments rather than relying solely on traditional, widely used response surface modeling approaches.

\begin{figure}[hbt!]
    \centering
    \includegraphics[width=1\textwidth]{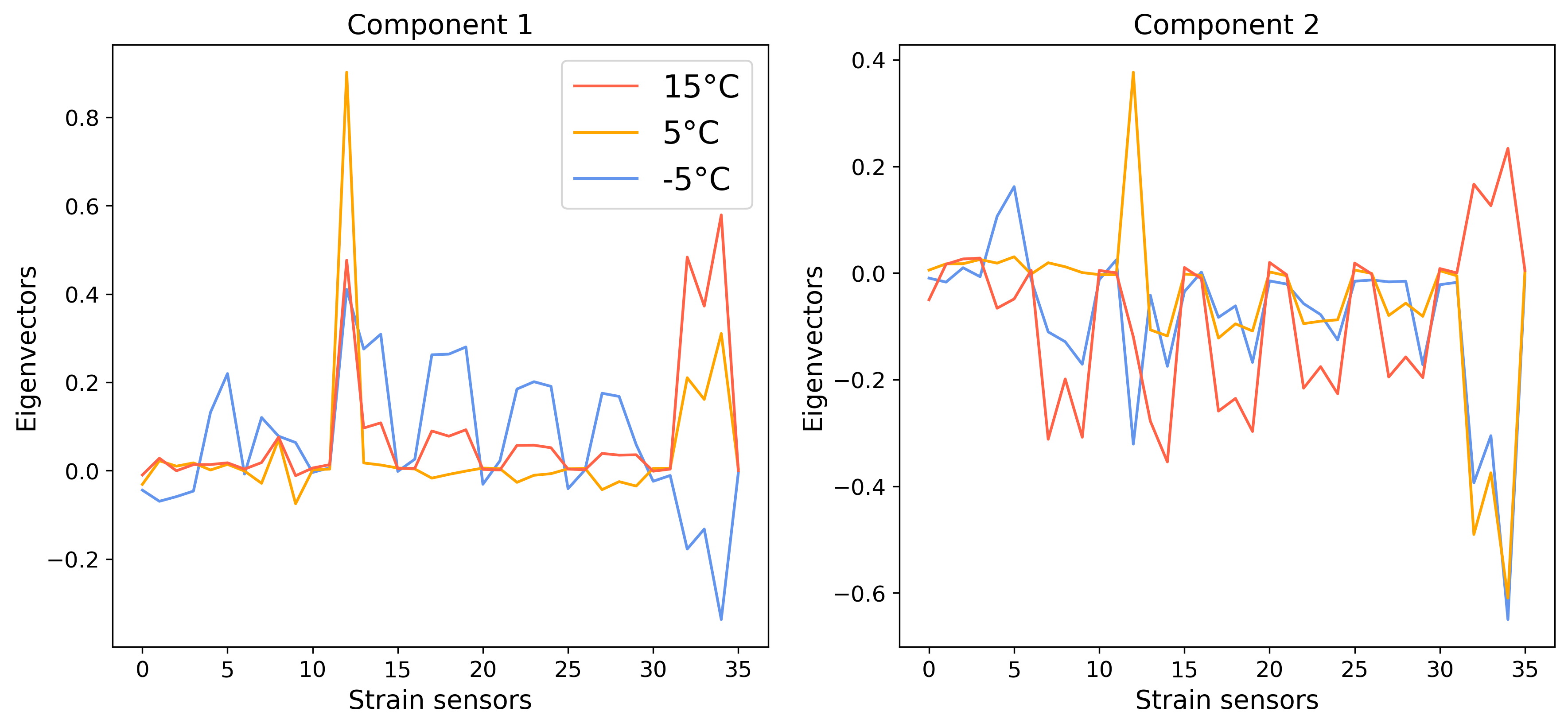}
    \caption{Spatial distribution of the first two principal components across the 36 strain sensors for three representative, equidistant temperatures (-5 °C, 5 °C, and 15 °C, respectively). Each curve shows the eigenvector components associated with the respective principal component.}
    \label{fig:eigenvectors_comp1+2.png}
\end{figure}

To further investigate how the temperature-dependent changes in the covariance structure manifest in the dominant modes of variation, a conditional principal component analysis was performed based on the estimated covariance matrices, see Section~\ref{Conditional Principal Component Analysis} for reference. 
Figure~\ref{fig:eigenvectors_comp1+2.png} shows the first two components in terms of the first two eigenvectors for three representative temperatures ($-$5~°C, 5~°C, and 15~°C).
Each curve represents the spatial distribution of the loadings across the 36 strain sensors. The x-axis corresponds to the sensor index. The figure illustrates how the dominant strain patterns vary across the sensor array at different temperatures. The spatial patterns for both the first and the second principal components can be seen to vary systematically with temperature across the 36 strain sensors. 
These observations further underscore temperature as a strong confounding factor in SHM, affecting not only the mean strain signals but also the covariance structure.

\begin{figure}[hbt!]
    \centering
    \includegraphics[width=1\textwidth]{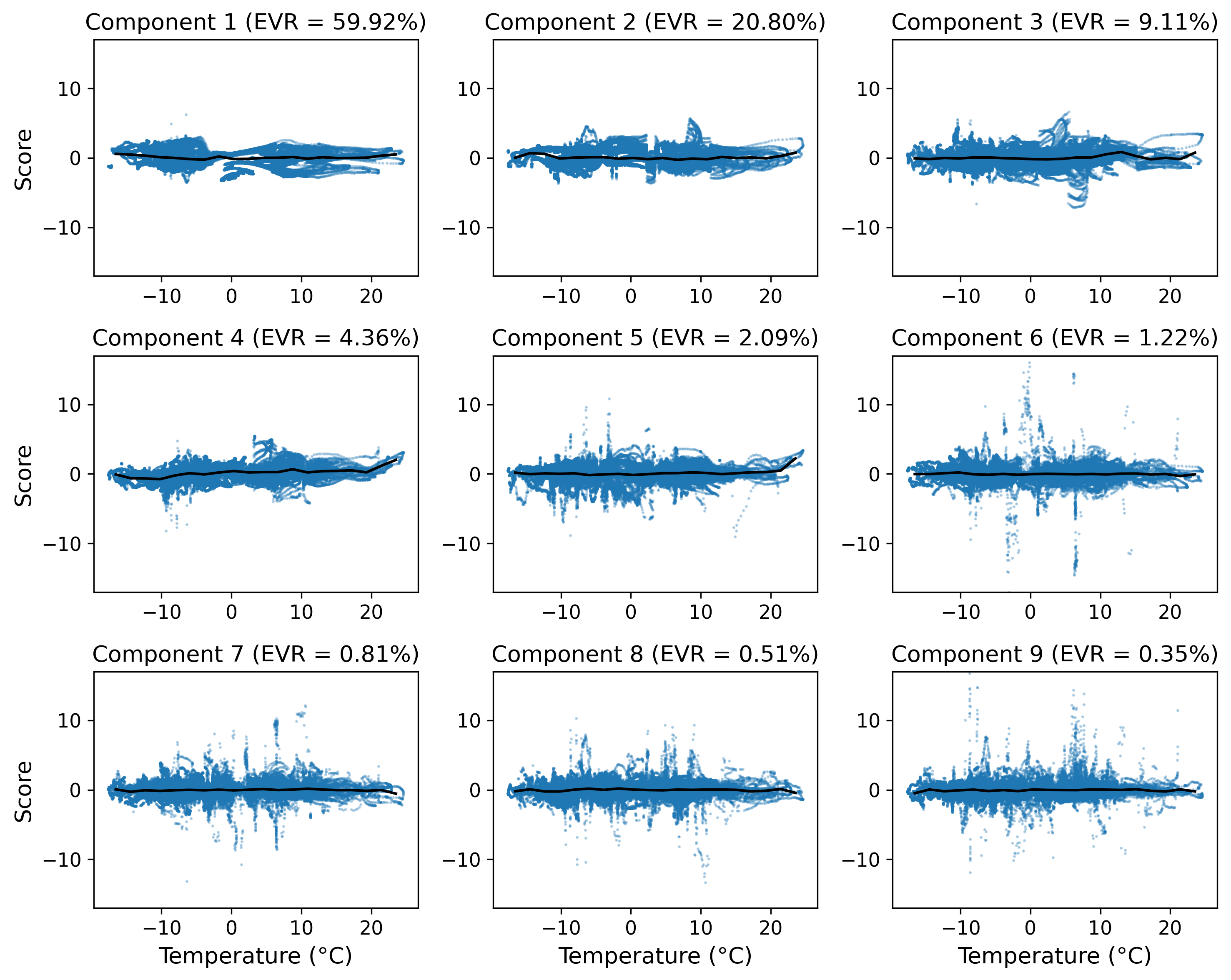}
    \caption{Scatter plots (blue) of selected conditional principal component scores (first nine components) against temperature, together with a nonparametric smoother (black) indicating the mean values as a function of the temperature.}
    \label{fig:scores_subplot.png}
\end{figure}

Proving the successful application of the methodology to the data, the scores (calculated as shown in (\ref{eq:scores})) are depicted against the temperature as scatter plots for the first 9 (of 36) components. This is shown in Figure \ref{fig:scores_subplot.png}. The black lines are non-linear (Nadaraya-Watson-type) smoothers. These mostly horizontal lines indicate a strongly reduced temperature influence across all components.

The explained variance ratio for component $j$ ($\mathrm{EVR}_j$) is defined via the variance of component~$j$ divided by the sum of variances of all components. In order to analyze this component-wise explained variance ratio, we first calculate temperature-specific $\mathrm{EVR}_j$ on a (fine) grid $\{z_1,\ldots,z_M\}$ of reasonable temperature-values via  

\begin{equation}
    \mathrm{EVR}_j(z_m)=\frac{\hat{\lambda}_j(z_m)}{\sum_{k=1}^p\hat{\lambda}_k(z_m)},
\end{equation}

with $p$ being the total number of components (36), and $\hat{\lambda}_j(z_m)$ the estimated eigenvalue of component $j$ at temperature $z_m$. Then, the average is computed in terms of

\begin{equation}
    \overline{\mathrm{EVR}}_j=\frac{1}{M}\sum_{m=1}^M\mathrm{EVR}_j(z_m).
\end{equation}

On the MOSAiC data considered, component 1 explains (on average) 59.92\% of the total variability, with components 2 and 3 further explaining 20.80\% and 9.11\%, respectively. Cumulatively, this adds up to around 90\% for the first three components and to a total of more than 99\% of the variability explained by the first nine components, as shown in Figure \ref{fig:scores_subplot.png}.

\section{Discussion}

In conclusion, the results of this work suggest that temperature exerts a substantial influence beyond the mean behavior of strain measurements and also on their covariance structure. This reinforces the central premise that environmental and operational variability affects multiple aspects of the statistical distribution of SHM data and cannot be sufficiently addressed by approaches that focus solely on first-order effects. In particular, the persistence of temperature-dependent correlation patterns even after detrending highlights the limitations of conventional response surface modeling. This has important implications for damage detection, as changes in dependency structures may be misinterpreted as structural anomalies if environmental effects are not properly accounted for.

Application of nonparametric conditional covariance estimation captures higher-order dependencies without imposing restrictive parametric assumptions. This work shows the successful application of the methodology to the icebreaker ``Polarstern''. Beyond the difference between static structures, such as bridges, and mobile structures, such as a maritime vessel, the MOSAiC expedition's Arctic conditions also served as an endurance test for the methodology, and the analysis of the data suggests similarly good application potential.

A crucial methodological aspect concerns the choice of the kernel bandwidth parameter $h$ in the covariance estimation. The bandwidth controls the degree of smoothing across temperature levels. Smaller values allow for capturing more localized variations, yet may increase estimation variability, particularly in regions with sparse data. Conversely, larger bandwidths yield smoother estimates but may obscure certain relevant structural changes. In this study, tests have been performed using multiple values of $h$, ultimately settling on $h=2$. The observed temperature-dependent patterns were found to be qualitatively robust across the range of bandwidth choices, suggesting that the main conclusions are not sensitive to the exact choice of bandwidth. Nevertheless, systematic bandwidth selection strategies, such as cross-validation, may further improve estimation accuracy in practical applications.

The presented methodology offers a promising direction for improving the reliability of SHM and the associated damage detection by accounting for complex environmental influences -- in this study, particularly highlighted in challenging settings such as long-term monitoring of marine structures under extreme conditions.

\bibliographystyle{unsrtnat}

\end{document}